\documentclass[11pt,twoside]{article}

\usepackage{asp2006}
\usepackage{epsf}
\usepackage{psfig}
\usepackage{lscape}

\def\gtrsim{\mathrel{\hbox{\rlap{\hbox{\lower4pt\hbox{$\sim$}}}\hbox{$>$}}}}

\begin{document}
\title{Parsec-Scale Jet-Environment Interactions in AGN}   
\author{Matthew L. Lister}   
\affil{Department of Physics, Purdue University}    

\begin{abstract} 

Observations made with the VLBA have led to fundamental advances in
our understanding of how radio jets in AGN evolve from parsec-scales
out to distances exceeding several hundred kiloparsecs. In this review
I discuss current models of young radio source evolution, as well as
the observational evidence for a rapid change in jet properties on
scales of $\sim 1$ kpc. A central topic of current debate is the
relative importance of intermittent jet fueling versus jet-environment
interactions in causing a drop-off in powerful radio sources at this
critical evolutionary stage. Recent 3-D hydrodynamical jet simulations
suggest that dense environments and cloud collisions can temporarily
stifle, but not completely halt powerful relativistic jets. Several
VLBA studies of jet-ISM interactions in both blazars and weak
Seyfert jets have indicated that collimated outflows are indeed
possible in dense environments. At present, the bulk of the evidence
favors intermittent AGN accretion as the dominant factor in
determining the evolutionary path of large numbers of AGN jets.

\end{abstract}


\section{Introduction}   

One of the most fundamental questions that can be asked about jets
associated with active galactic nuclei (AGN) is how do they evolve from
their dense, gas-rich parsec-scale environments out to scales of
hundreds of kiloparsecs, well outside their host galaxies.  The
capability of radio wavelength interferometers to penetrate the dense
gas and dust in the centers of AGN host galaxies at high resolution
has brought us tantalizingly close to fully answering this
question. In this review, I briefly describe our current understanding
of young radio jet evolution, and the
relative role played by jet-environment interactions. I begin in
\S2 by discussing what has been learned from statistical
population studies, and devote Sections~3 and 4 to numerical jet
simulations and individual VLBA case studies that have improved our
understanding of interactions between AGN jets and their parsec-scale
environments.

\section{\label{evolution}Evolution of young AGN jets}

Our current knowledge of radio jet evolution owes a great deal to the
gigahertz-peaked spectrum (GPS) class of radio source, which comprise
approximately $\sim 10\%$ of flux-limited samples at
cm-wavelengths. Originally classified in early surveys as 'compact
doubles' by \cite{PM82}, subsequent improvements in VLBI capabilities
revealed weak central components, and in some cases faint bridges of
emission connecting them with bright outer features. It was soon
recognized that these AGN were miniature versions of the classical
kpc-scale lobe-core-lobe radio galaxies, with similar total radio
powers, but over a thousand times smaller in extent.

Based on observed size trends in GPS and compact steep spectrum (CSS)
sources (e.g., \citealt*{JS02}), self-similar expansion models (e.g.,
\citealt*{Beg96,BDO97}) were developed in which the overall linear extent
of the jets grow in proportion with their hotspot diameters. These
hotspots remain in ram pressure equilibrium with the external medium,
which implies that the evolution of the source is strongly dictated by
the density profile of the ISM. Numerical simulations (see \S~3) of
jets expanding into power-law external density profiles confirmed that
a large bow shock forms ahead of the hotspot, allowing the latter to
expand smoothly and propagate outward relatively unimpeded. Unlike the
dentist drill model for kpc-scale lobes, very little side-to-side
motion is expected for the pc-scale hotspot. Spectacular confirmation
of these models came with the first measurements of hotspot proper
motions in GPS radio galaxies (e.g., \citealt*{OCP99}), which
displayed predominantly outward (non-transverse) motion. The derived
kinematic ages, based on constant expansion, were typically $\sim
1000$ y \citep{GTP05}, confirming that these were in fact recently
launched jets.

The first problems with the standard scenario arose with detailed
studies of population statistics. In a steady-state population, one
would expect a rather flat distribution of kinematic ages, but in
fact, the observed one is peaked at young ages
\citep{GTP05}. A similar conclusion had been reached previously 
by independent authors who considered the luminosity functions of GPS
sources (e.g., \citealt*{Beg96,RTP96}). Given their high luminosities,
the young radio sources were too numerous compared to their more aged
radio galaxy cousins, implying that must either dim rapidly, or die
out completely before reaching sizes of a few kpc. A lingering issue
of current debate is the relative importance of AGN fueling and
environmental interactions in dictating the evolution of radio jets at
this critical evolutionary stage.

\subsection{AGN fueling and intermittent jet activity}

Although a simple argument for intermittent jet activity in AGN can be
found in the fact that only $\sim10\%$ of all AGN associated with
super-massive black holes are radio loud, yet the lifetimes of
individual AGN are on the order of a few hundred Myr, true 'smoking
gun'-type evidence has become available only relatively recently. The
most compelling has been the discovery of the 'double-double' class of
radio galaxy \citep{SDR00}, of which roughly a dozen are currently
known \citep{MTM06}. These sources contain two sets of nested radio
lobes, which are symmetric with respect to a central component
associated with the active nucleus. The inner double resembles in many
ways a GPS source, with a peaked radio spectrum, bright hotspots, and
fast expansion speed. The outer lobe structures, on the other hand,
have sizes comparable to the those of the largest known radio
galaxies.  The notable gap in radio emission between the two
components is indicative of a long quiescent period, on the order of
$10^6-10^7$ Myr, in which the jet was presumably switched off (e.g.,
\citealt*{OKB01}).

Understanding intermittent jet activity in AGN is undoubtedly an
important factor in building a complete model of jet evolution (e.g.,
\citealt*{RB97}). However, it is still a nascent field in which the
necessary statistical samples (needed because of the long evolutionary
timescales involved) are still being gathered.  As I will describe
below, considerably larger progress has been made in understanding the
role played by jet-ISM interactions in affecting AGN jet evolution.

\subsection{Basic forms of jet-ISM interaction}

Because they are relatively light compared to their external
environments (density contrasts on the order of $10^{-3}$,
e.g., \citealt*{Krause03}), AGN jets are highly susceptible to external
interactions, which can be classified roughly into three main areas:
\begin{itemize}

\item      {\bf Bow shock-hotspot interaction} at the jet terminus, as
in the standard models described above. 

\item       {\bf Cloud collisions}, which can cause bending and disruption of the flow.

\item        {\bf Entrainment}, leading to shear layers, deceleration,
instabilities, and possible particle acceleration at the jet boundaries.
\end{itemize}

Although much is known about the physics of entrainment in
kiloparsec-scale jets, progress on parsec-scales has been limited by
several factors. These include the difficulty of observing faint,
diffuse emission at the jet boundaries with limited dynamic-range
VLBI, as well as a paucity of bright, nearby AGN jets which we can
resolve in a transverse direction to the flow. Furthermore, studies of
the crucial 100-1000 milliarcsecond region where jets may undergo
strong internal changes due to entrainment have been hampered by the
lack of a suitable interferometer matching the sensitivity of the VLA
or VLBA. For these reasons I will concentrate hereafter on the issue
of jet interactions with dense clouds in the nuclear region of the
host galaxy.

\section{\label{sims}Numerical jet-cloud simulations}

Numerical simulations continue to play a vital role in understanding
the structure and evolution of AGN jets, by providing the ability to
test various scenarios under controlled conditions. Early numerical
jet-medium interaction studies were able to reproduce classical bow
shock and hotspot structures by propagating supersonic outflows into
external media with uniform density and pressure gradients (e.g.,
\citealt*{HN90}). The extension of MHD codes to the fully
three-dimensional, relativistic regime has made it possible to
robustly examine powerful jet evolution through a more realistic,
non-uniform medium for the first time. I describe here two such
studies (\citealt{CW07}, and \citealt*{SB07}), that are of particular
relevance to young jet evolution.

The simulations of \cite{CW07} employ a fully 3-D, pure hydrodynamic
code to simulate the passage of the relativistic jet through a
two-phase medium. The latter consists of a single dense cloud embedded
in a constant-pressure gas. They examined cases of both high ($\Gamma
= 7$) and low ($\Gamma = 2.29$) Lorentz factor jets striking the cloud
slightly off-axis. During the interaction, an oblique shock forms in
the jet, causing it to bend. Unlike previous non-relativistic studies
(e.g., \citealt*{wang00,higgins99}), the flow itself does not undergo
any significant deceleration or decollimation, and remains stable
after the interaction event. By varying the cloud-to-ambient medium
ratio, the authors find that the highest deflections occur in the case
of low-Mach number jets hitting denser clouds, with cloud density
being the dominant factor. Thicker clouds end up being less
encompassed by the bow shock, allowing earlier interaction with the
Mach disk and stronger oblique shocks in the flow. The clouds
themselves can actually survive the event, provided the cloud/jet
density contrast is high enough to suppress most Kelvin-Helmholtz
instabilities. These regions of shocked gas may be important star
formation sites (see \S~4) and may play a role in creating the
emission-line/jet alignment effect in AGN (e.g.,
\citealt*{MBS87}).

\cite{SB07} investigate the more general case of a
jet propagating through an inhomogeneous medium in the form of a
massive ($10^{10} \;\mathrm{M_{\sun}}$), turbulently supported disk
plus a hot ($10^7$ K) ISM. Like \cite{CW07}, they use a fully 3-D pure
hydrodynamic code, although in this case a non-relativistic one for
which they derive relativistic scaling parameters according to
\cite{KF96}.  In the initial phase of their simulations of a $\sim
10^{43}\; \mathrm {erg\; s^{-1}}$ jet, the morphology looks strikingly
different than those seen in other studies that assume a uniform ISM,
in that the flow attempts to seek out and pass through the
lowest-density locations in the clumpy (fractal) medium. In doing so,
multiple channels are formed and reformed, followed by the formation
of quasi-spherical bubbles around the jet and counter-jets that expand
outward. Making simple assumptions about the gas emissivity, the
authors find that these bubbles should be prominent in hard
X-rays. Once the jet reaches the outer edge of the disk and clears the
last obstruction, a stable, linear outflow develops, containing the
standard re-collimation and bow shock structures. At this point it
pierces the expanding bubble and evolves as in the uniform medium
case.

The authors find a good deal of similarity between the predicted radio
emission from their simulations and the compact symmetric object (CSO)
4C~31.04 \citep{CFG95}. This young radio source is characterized by a
large asymmetry in its jet and counter-jet structure, as well as lobe
spectral index gradients that are difficult to reconcile with standard
models of cocoon backflow \citep{giro03}. Comparison with their
simulations led \cite{SB07} to suggest that the western lobe may be
near the end of the breakout phase, whereas the eastern lobe is at a
slightly earlier stage of evolution. The strong apparent northward
deflection of the western lobe flow at the hotspot is also reminiscent
of structure found in the simulations of \cite{CW07}.

The conclusion that can be drawn from these studies is that powerful
relativistic jets are not likely to be permanently stifled by neither
direct jet-cloud collisions, nor a dense, clumpy external
medium. Instead, it is more likely that they all pass through an
evolutionary stage in which the flow may be bent and not necessarily
well-collimated. The duration of this stage is largely determined by
the power of the jet, and to a lesser extent, the jet/medium density
contrast. Given the good initial agreement with observed jet structure
from these preliminary simulations, it suggests that through careful
study of jet morphologies of young radio sources, it may be possible
to identify the precursors to both high- and low-power radio
galaxies, as well as to characterize their early evolutionary paths.

\section{\label{cases}VLBA studies of jet-environment interactions}

In addition to providing measurements of kinematic expansion speeds,
the VLBA provides a variety of unique tools for studying jet-medium
interactions on parsec scales. These include HI absorption
measurements, Faraday de-polarization and electric vector rotation
measurements at sub-milliarcsecond resolution levels. I discuss here
several recent VLBA studies of ISM interactions in weak Seyfert
jets, as well as in powerful blazars.

\subsection{Seyfert galaxies}

The relative proximity (15-20 Mpc) of Seyfert galaxies makes them
ideal targets for investigating jet-environment effects with the VLBA
at spatial resolutions approaching several thousand A.U. Given that
their jet powers are typically a factor of 100-1000 smaller than
radio-loud quasars (e.g., \citealt*{gold99}), they are much more
subject to entrainment and disruption (e.g.,
\citealt*{deyoung06}). Their sporadic accretion rate also offers the
chance to examine in detail the effects of central engine disruption
on jet structure.

\subsubsection{NGC 4151:}
The nearly-face on Seyfert 1.5 galaxy NGC 4151 has been the subject of
many intensive VLBI studies, due to its well-defined, two-sided,
$\sim100$ parsec-long radio jets, as well as the large quantity of
neutral gas in its nuclear region. HST imaging has revealed numerous
ionized gas clouds in an inner region that is extended about an axis
roughly aligned with the radio jets \citep{hutchings98, kaiser00}. The
spatial geometry of the narrow-line region suggests a thick molecular
torus aligned perpendicular to the jet, which is confirmed by
$\mathrm{H_2}$ measurements \citep{fernan99}. VLBA absorption data
have also provided evidence for an inner HI ring \citep{ulvestad98,
mundell03}. The radio spectral flattening and brightness enhancement
of the jet at this location led \cite{mundell03} to suggest that this
marks a site of jet-ISM interaction. Although the VLBA images lack
sufficient dynamic range to fully examine the extremely weak surface
brightness structure, the jet does undergo an abrupt deviation at this
point, in a manner similar to the jet-cloud simulations of
\cite{CW07}. \cite{mundell03} found the HI absorption line profiles to
vary significantly toward different portions of the jet, indicating a
medium composed of clumpy dense clouds with a variety of
velocities. Although they speculate that some of the other bright
knots in the jet may be the result of jet-cloud encounters, the
authors rule out shock ionization as the main source of the NLR, based
on its imprecise alignment with respect to the radio jet, and the
presence of several low-velocity clouds very near the jet that show no
signs of interaction. 

\subsubsection{NGC 3079:}

This is another good example of a Seyfert jet in a dense environment,
albeit in this case the galaxy is viewed nearly edge-on
\citep{sosa01}. Using a series of VLBA measurements over a six year
period, \cite{middel07} have discovered complex kinematics and
variable jet emission in this source. They found one bright jet knot
initially moving at nearly 0.1 c, only to watch it decelerate and
become virtually stationary during the final year of their
observations. During this time its flux density increased and its
spectrum changed to a convex free-free/synchrotron-self absorbed
profile. This behavior is consistent with that expected from the
jet-cloud simulations described in \S~\ref{sims} Furthermore, the
source contains several steeper spectrum features well off the main
jet axis, which could perhaps be remnants of earlier flow channels
as predicted by \citealt*{SB07}. NGC 3079 thus provides an excellent example
of the potential of multi-epoch VLBA studies for exploring the
kinematics of jet-cloud interactions at exceedingly high spatial
resolution.

\subsubsection{PKS 1345+12:}

The ultra-luminous infrared galaxy IRAS 13451+1232 is a recent merger
system with significantly distorted optical morphology and a binary
nucleus, the northwest of which has been classified as a Seyfert 2
(e.g., \citealt*{scoville00}). The latter also contains a spectacular
radio jet (PKS 1345+12), which extends nearly 200 pc in a continuous,
sinusoidal pattern \citep{LKV03}. The counter-jet is also visible, but
only out to $\sim 50$ pc from the nucleus. Although these properties
are consistent with the CSO class, this object is unique in the fact
that \cite{LKV03} measured speeds of 1 c in the innermost
jet region, as well as high fractional polarization at the location of
the southern hotspot. The latter is significant as it implies a
continuous resupply of energy, i.e., the southern jet is not stifled
by this very gas rich galaxy.

By fitting to the apparent ridge line, apparent speeds, and
jet/counter-jet ratio, \cite{LKV03} concluded that the jet follows a
three-dimensional, conical helix aligned 82 degrees from our line of
sight, with an intrinsic flow speed of $\sim 0.8$ c. Similar sinusoidal
ridge lines seen in other CSOs and blazars have led various authors to
conclude that these may be the result of growing Kelvin-Helmholtz
instability modes, driven by small perturbations at the jet nozzle and
excited by interaction with the medium at the jet boundaries. The
northern counter-jet shows a deviation from the predicted best-fit
helical path, and is truncated at the site of dense HI absorption
($>10^{22} \;\mathrm{cm^{-2}}$; \citealt*{morganti05}). This appears
therefore to be a clear case where asymmetries in the external
environment have a strong differential impact on the morphology and
evolutionary rates of the jet and counter-jet of a young radio source.

\subsection{Blazar Jets}

Despite their much larger distances, blazar jets can also serve as
useful probes of parsec-scale jet interactions. First, because they
are viewed directly down the opening in the obscuring torus, there is
much less de-polarization, meaning that the jet polarization and
magnetic field properties can be directly studied. This also means
that any intervening gas can be potentially studied via Faraday
rotation measures (e.g., \citealt*{ZT05}). Second, any slight
deviations in the flow that may be caused by interactions are greatly
magnified by projection effects. Finally, because of Doppler effects,
there are many examples of blazars where over a century of jet evolution is
compressed into a span of only a few years of observing time (e.g.,
\citealt*{KL04}). 

\subsubsection{3C 279:}

The powerful jet in the quasar 3C279 was one of the first jets in
which superluminal motion was witnessed, and has been the target of
intensive study in a variety of wave-bands. The jet has been
regularly imaged since 1994 by the 2 cm Survey \citep{KL04} and MOJAVE
\citep{LH05} programs with the VLBA at a wavelength of 2 cm. Shorter
wavelength (7 mm) VLBA monitoring \citep{jorstad04,jorstad05} has
revealed a regular swing in the ejection direction of the jet close to
the nozzle, over a timescale of 3 years. \cite{homan03} describe one
prominent jet feature (C4) that was ejected in late 1984, which moved
steadily along a linear path for over a decade with an apparent speed
of 8 c, before suddenly undergoing an increase in brightness and
change in polarization angle in 1998. These events were followed
shortly thereafter by a rapid apparent acceleration to 13 c, and a
change in trajectory of 26 degrees. Under the most conservative
assumptions, \cite{homan03} found that these changes were consistent
with an intrinsic bend of only 0.5 to 1 degree. Given the fact the
brightening and polarization changed {\it before} the change in
trajectory, the most plausible scenario is one in which C4 is
interacting with the external environment. Furthermore, the direction
of the new trajectory closely matches that of another feature ejected
several decades previously, which rules out a random jet-cloud
collision. The authors suggest instead that the event represents a
collimation of the jet resulting from a jet-boundary interaction at a
de-projected distance $\gtrsim 1$ kpc from the nucleus. Since this is
the first such an event to be witnessed in an AGN jet, it is difficult
to yet draw solid conclusions on the validity of this model. However,
large intensive VLBA monitoring programs such as MOJAVE \citep{LH05}
may soon provide additional examples for further study.

\subsubsection{3C 120:}

Although classified as a Seyfert 1, this nearby (z = 0.033)
broad-lined galaxy shares many properties with blazars, including
superluminal motions of up to 6 c, a one-sided radio jet, and flux
variability. \cite{axon89} found high-velocity emission line
components in the host galaxy that suggested interaction between the
jet and gas clouds in the NLR. The excellent spatial resolution (0.1
pc) achievable by the VLBA at 43 GHz has enabled detailed study of its
jet evolution in both total intensity and linear polarization
\citep{gomez01, jorstad05}. The jet is resolved perpendicular to the
flow direction, and a distinct asymmetry is seen between the northern
and southern edges. In particular, \cite{gomez01} have found a
distinct region in the southern edge, approximately 8 pc
(de-projected) from the base of the jet, where moving jet features
show marked changes as they pass through. These include a brightening
in flux density, and a rotation of their polarization electric vector
position angles (EVPAs). These events are different from that
witnessed in 3C 279, since in this case no accelerations are
seen. \cite{gomez01} conclude the most likely explanation to be
interaction with a cloud, which causes Faraday rotation of the EVPAs,
and shocking of the jet material. There is also an indication of a
slight bend at the interaction site, although the jet remains
well-collimated downstream. Ideally it would be useful to study
additional examples of this type of interaction, but unfortunately
there are still very few known bright jets that are close enough to be
resolved transversely by the VLBA, and yet have viewing angles small
enough not to be heavily de-polarized by foreground nuclear gas.

\section{Summary}

High-resolution radio observations of young radio jets associated with
gigahertz-peaked spectrum AGN have led to considerable insight into
the evolutionary processes of AGN jets. Kinematic and population
studies have shown that these young radio sources undergo a
significant decline in numbers when they reach sizes of $\sim 1$
kpc. VLBA studies of individual jets have provided clear evidence for
interaction with clouds in their external environment, suggesting
stifling by dense gas as a possible cause. However, detailed
numerical simulations of jet-environment interactions indicate that
dense, clumpy environments can only temporarily stifle the flow of
powerful jets, even in the case of direct jet-cloud
collisions. Furthermore, the discovery of 'double-double' galaxies has
provided solid evidence of recurrent jet activity in powerful AGN. It
therefore appears likely that variable accretion rates play a major
role in determining the evolutionary paths of many AGN. The enhanced
resolution and sensitivity of upcoming facilities such as VSOP-II, the
EVLA, and the SKA should provide many new opportunities for studying
the evolution of young radio sources and their interactions with their
external environment.


\begin{thebibliography}{}

\bibitem[Axon et al.(1989)]{axon89} Axon, D.~J. et al. 1989, \nat, 341, 631 

\bibitem[Begelman(1996)]{Beg96} Begelman, M. C. 1996, in Study of a Radio Galaxy, ed. C. Carilli \& D. Harris (Cambridge: Cambridge Univ. Press), 209 (B96)

\bibitem[Bicknell et al.(1997)]{BDO97} Bicknell, G.~V., Dopita, M.~A., \& O'Dea, C.~P.~O.\ 1997, \apj, 485, 112 

\bibitem[Choi et al.(2007)]{CW07} Choi, E., Wiita, P. J., Ryu, D. 2007, \apj, 655, 769

\bibitem[Cotton et al.(1995)]{CFG95} Cotton, W.~D. et al. 1995, \apj, 452, 605 

\bibitem[De Young(2006)]{deyoung06} De Young, D.\ 2006, Astronomische Nachrichten, 327, 231 

\bibitem[Fernandez(1999)]{fernan99} Fernandez, B. R. et al. 1999, \mnras, 305, 319

\bibitem[Giroletti et al.(2003)]{giro03} Giroletti, M.  et al. 2003, \aap, 399, 889

\bibitem[Goldschmidt et al.(1999)]{gold99} Goldschmidt, P. et al. 1999, \apj, 511, 612 

\bibitem[Gomez et al.(2001)]{gomez01} Gomez, J.-L. et al. 2001, \apj, 561, L161

\bibitem[Gugliucci et al.(2005)]{GTP05} Gugliucci, N.~E., Taylor, G.~B., Peck, A.~B., \& Giroletti, M.\ 2005, \apj, 622, 136 

\bibitem[Hardee \& Norman(1990)]{HN90} Hardee, P. E. \& Norman M. L. 1990, \apj, 365, 134

\bibitem[Higgins et al.(1999)]{higgins99} Higgins, S.~W., 
O'Brien, T.~J., \& Dunlop, J.~S.\ 1999, \mnras, 309, 273 

\bibitem[Homan et al.(2003)]{homan03} Homan et al. 2003, \apj, 589, L9

\bibitem[Hutchings et al.(1998)]{hutchings98} Hutchings, J. B. et al. 1998, \apj, 492, L115

\bibitem[Jeyakumar \& Saikia(2002)]{JS02} Jeyakumar, S., \& 
Saikia, D.~J.\ 2002, New Astronomy Review, 46, 421 

\bibitem[Jorstad et al.(2004)]{jorstad04} Jorstad, S. et al. 2004, \aj, 127, 3115

\bibitem[Jorstad et al.(2005)]{jorstad05} Jorstad, S. et al. 2005, \aj, 130, 1418

\bibitem[Kaiser et al.(2000)]{kaiser00} Kaiser, M. E. et al. 2000, \apj, 528, 260

\bibitem[Kellermann et al.(2004)]{KL04} Kellermann et al., 2004, \apj, 609, 539

\bibitem[King \& Pringle(2007)]{2007MNRAS.377L..25K} King, A.~R., \& Pringle, J.~E.\ 2007, \mnras, 377, L25 

\bibitem[Kommissarov \& Falle(1996)]{KF96} Kommissarov, S. S. \& Falle, S. A. E. 1996, in ASP Conf. Series, 173

\bibitem[Krause(2003)]{Krause03} Krause, M.\ 2003, \aap, 398, 113 

\bibitem[Lister et al.(2003)]{LKV03} Lister, M.~L. et al. 2003, \apj, 584, 135 

\bibitem[Lister \& Homan(2005)]{LH05} Lister, M.~L.,~\& Homan, D.C. 2005, \aj, 130, 1389

\bibitem[McCarthy et al.(1987)]{MBS87} McCarthy, P.~J., van 
Breugel, W., Spinrad, H., \& Djorgovski, S.\ 1987, \apjl, 321, L29

\bibitem[Marecki et al.(2006)]{MTM06} Marecki, A., Thomasson, 
P., Mack, K.-H., \& Kunert-Bajraszewska, M.\ 2006, \aap, 448, 479 

\bibitem[Middelberg et al.(2007)]{middel07} Middelberg, E. et al. 2007, \mnras, 377, 731

\bibitem[Morganti et al.(2005)]{morganti05} Morganti, R., Tadhunter, C.~N., \& Oosterloo, T.~A.\ 2005, \aap, 444, L9
 
\bibitem[Mundell et al.(2003)]{mundell03} Mundell, C. G. et al. 2003, \apj, 583, 192 

\bibitem[O'Dea et al.(2001)]{OKB01} O'Dea, C.~P., Koekemoer, 
A.~M., Baum, S.~A., Sparks, W.~B., Martel, A.~R., Allen, M.~G., Macchetto, F.~D., \& Miley, G.~K.\ 2001, \aj, 121, 1915 

\bibitem[Owsianik et al.(1999)]{OCP99} Owsianik, I., Conway, 
J.~E., \& Polatidis, A.~G.\ 1999, New Astronomy Review, 43, 669 

\bibitem[Phillips \& Mutel(1982)]{PM82} Phillips, R.~B., \& 
Mutel, R.~L.\ 1982, \aap, 106, 21 

\bibitem[Readhead et al.(1996)]{RTP96} Readhead, A.~C.~S., 
Taylor, G.~B., Pearson, T.~J., \& Wilkinson, P.~N.\ 1996, \apj, 460, 634 

\bibitem[Reynolds \& Begelman(1997)]{RB97} Reynolds, C.~S., 
\& Begelman, M.~C.\ 1997, \apjl, 487, L135 

\bibitem[Schoenmakers et al.(2000)]{SDR00} Schoenmakers, 
A.~P., de Bruyn, A.~G., R{\"o}ttgering, H.~J.~A., \& van der Laan, H.\ 
2000, \mnras, 315, 395 


\bibitem[Scoville et al.(2000)]{scoville00} Scoville, N. Z. et al. 2000, \aj, 119, 991

\bibitem[Sosa-Brito et al.(2001)]{sosa01} Sosa-Brito, R. M. et al. 2001, \apjs, 136, 61

\bibitem[Sutherland \& Bicknell(2007)]{SB07} Sutherland, R.S., \&
Bicknell, G. V. 2007, \apj, submitted

\bibitem[Ulvestad et al.(1998)]{ulvestad98} Ulvestad, J. S. et al. 1998, \apj, 496, 196

\bibitem[Wang et al.(2000)]{wang00} Wang, Z., Wiita, P. J., \& Hooda, J. S. 2000, \apj, 534, 201

\bibitem[Zavala \& Taylor(2005)]{ZT05} Zavala, R. T. \& Taylor, G. B. 2005, \apj, 612, 749

\end{thebibliography}

{}

\end{document}